\documentstyle[seceq]{ptptex}

\def\VEV#1{\left\langle #1\right\rangle}
\def\tr{\mathop{\rm tr}\nolimits}

\def\sqr#1#2{{\vcenter{\hrule height.#2pt
      \hbox{\vrule width.#2pt height#1pt \kern#1pt
          \vrule width.#2pt}
      \hrule height.#2pt}}}

\def\square{{\mathchoice{\sqr84}{\sqr84}{\sqr{5.0}3}{\sqr{3.5}3}}}
\def\dA{\mathop\square\nolimits}
\def\Mto{{\buildrel{M\rightarrow\infty}\over=}}
\preprintnumber[3cm]{
IU-MSTP/34\\ hep-th/9905225}

\title{
Invariant Regularization of Supersymmetric Chiral Gauge Theory%
\footnote{Based on the invited talk in the
workshop on ``Gauge Theory and Integrable Models,'' January 1999,
Kyoto}
}

\author{
Hiroshi {\sc Suzuki}\footnote{E-mail address:
hsuzuki@mito.ipc.ibaraki.ac.jp}
}

\inst{
Department of Physics, Ibaraki University, Mito 310-8512, Japan
}

\recdate{
\today
}

\abst{
We present a regularization scheme which respects the supersymmetry
and the maximal background gauge covariance in supersymmetric
{\it chiral\/} gauge theories. When the anomaly cancellation
condition is satisfied, the effective action in the superfield
background field method automatically restores the gauge invariance
without counterterms. The scheme also provides a background gauge
covariant definition of composite operators that is especially useful
in analyzing anomalies. We present several applications: The minimal
consistent gauge anomaly; the super-chiral anomaly and the
superconformal anomaly; as the corresponding anomalous commutators,
the Konishi anomaly and an anomalous supersymmetric transformation
law of the supercurrent (the ``central extension'' of
$N=1$~supersymmetry algebra) and of the $R$-current.}

\begin{document}

\maketitle

\section{Introduction}
Behind the recent non-perturbative analyses of the supersymmetric
gauge theories, the existence of an invariant regularization that
respects the supersymmetry and the gauge symmetry always seems to be
assumed. Such a regularization should be vital also in practical
calculations because it avoids the introduction of counterterms
necessary to recover Ward-Takahashi identities. Therefore, it
seems yet to be important to seek for such an invariant
regularization in supersymmetric gauge theories.\cite{rf:JAC} \
If one knows how to restore the preferred symmetries by counterterms,
of course, any regularization may be employed. However, it is also
true that with a non-invariant regularization a naive analysis easily
leads to wrong physical predictions in view of the preferred
symmetries. For example, one might miss the fermion number
non-conservation in chiral gauge theories,\cite{rf:THO} with use of
the gauge non-invariant Pauli-Villars regularization. An invariant
regularization allows one to obtain correct predictions in the naive
way.

In this talk, I like to introduce the recent attempt of our
group\cite{rf:HAY}\tocite{rf:OHS} for obtaining an invariant
regularization scheme in supersymmetric {\it chiral\/} gauge
theories. Our scheme is perturbative in nature but it possesses the
following properties. The supersymmetry is manifest; the scheme is
based on the superfield formalism in an exactly four-dimensional
spacetime. The effective action in the superfield background field
method {\it above\/} one-loop order is always background gauge
invariant. Being consistent with the existence of the gauge anomaly,
the one-loop effective action is not background gauge invariant in
general. However the breaking is kept to be minimal; the consistent
gauge anomaly is proportional to the
anomaly~$d^{abc}=\tr T^a\{T^b,T^c\}$. As the result, when the anomaly
cancellation condition~$d^{abc}=0$ is satisfied, the effective action
restores the background gauge invariance without any counterterms. The
scheme also provides a background gauge covariant supersymmetric
definition of composite operators, that is especially useful in
analyzing anomalies. Unfortunately the quantum gauge (or BRS)
symmetry is not manifest in our scheme and one has to verify the
Slavnov-Taylor identity order by order (this is possible at least
in the one-loop level\cite{rf:HAYA}). We note that however even with
respect to the background gauge symmetry it is not trivial to obtain
such an invariant scheme, due to the gauge anomaly in chiral gauge
theories.

Our convention is basically that of Ref.~\citen{rf:WES}; our
particular convention will be noted when necessary. For simplicity of
presentation, we assume the gauge representation~$R$ of the chiral
multiplet, which will be denoted as~$T^a$, is irreducible. The
normalization of the gauge generator is $[T^a,T^b]=it^{abc}T^c$,
$\tr T^aT^b=T(R)\delta^{ab}$,
$(T^a)_{ij}(T^a)_{jk}=C(R)\delta_{ik}$ and
$t^{acd}t^{bcd}=C_2(G)\delta^{ab}$.

%
%
%

\section{Superfield background field method}
We consider a general renormalizable supersymmetric
model:\footnote{%
$d^8z=d^4x\,d^2\theta\,d^2\overline\theta$, $d^6z=d^4x\,d^2\theta$,
$d^6\overline z=d^4x\,d^2\overline\theta$.}
\begin{equation}
   S={1\over2T(R)}\int d^6z\,\tr W^\alpha W_\alpha
   +\int d^8z\,\Phi^\dagger e^V\Phi
   +\int d^6z\,\Bigl({1\over2}\Phi^T m\Phi+{1\over3}g\Phi^3\Bigr)
   +{\rm h.c.}
\label{eq:twoxone}
\end{equation}
To apply the notion of the superfield background field
method,\cite{rf:GRI} we split the gauge superfield and the chiral
superfield as\cite{rf:HAY}
\begin{equation}
   e^V=e^{V_B}e^{V_Q},\qquad\Phi=\Phi_B+\Phi_Q.
\label{eq:twoxtwo}
\end{equation}
Here, the subscripts $B$ and~$Q$ represent the background field and
the quantum field, respectively. We shall regard $V_B$~as a vector
superfield, and thus $V_Q$~is {\it not\/} a vector superfield:
$V_Q^\dagger=e^{V_B}V_Q e^{-V_B}$. With the
splitting~(\ref{eq:twoxtwo}), the original gauge
transformation\cite{rf:WES}
\begin{equation}
   e^{V'}=e^{-i\Lambda^\dagger}e^Ve^{i\Lambda},
   \qquad\Phi'=e^{-i\Lambda}\Phi,
\label{eq:twoxthree}
\end{equation}
where $\Lambda=T^a\Lambda^a$ is a chiral superfield
$\overline D_{\dot\alpha}\Lambda=0$, is realized in the following two
different ways. (i) By the quantum gauge transformation:
\begin{equation}
   V_B'=V_B,
   \qquad e^{V_Q'}
   =(e^{-V_B}e^{-i\Lambda^\dagger}e^{V_B})e^{V_Q}
   e^{i\Lambda},
   \qquad\Phi'=e^{-i\Lambda}\Phi.
\label{eq:twoxfour}
\end{equation}
(ii) By the background gauge transformation:
\begin{equation}
   e^{V_B'}=e^{-i\Lambda^\dagger}e^{V_B}e^{i\Lambda},
   \qquad V_Q'=e^{-i\Lambda}V_Qe^{i\Lambda},
   \qquad\Phi'=e^{-i\Lambda}\Phi.
\label{eq:twoxfive}
\end{equation}
Next, we introduce the background covariant derivative symbol:
\begin{equation}
   \nabla_\alpha=e^{-V_B}D_\alpha e^{V_B},\qquad
   \overline D_{\dot\alpha},\qquad
   \{\nabla_\alpha,\overline D_{\dot\alpha}\}
   =-2i\sigma^m_{\alpha\dot\alpha}\nabla_m.
\label{eq:twoxsix}
\end{equation}
Since the gauge parameter~$\Lambda$ in~Eq.~(\ref{eq:twoxfive}) is
chiral, these derivatives transform as
$\nabla'=e^{-i\Lambda}\nabla e^{i\Lambda}$ under the background field
transformation~(\ref{eq:twoxfive}). On the gauge representation~$R$,
the covariant derivative is defined by
\begin{equation}
   {\cal D}_\alpha\Phi=\nabla_\alpha\Phi,
   \qquad{\cal D}_m\Phi=\nabla_m\Phi.
\label{eq:twoxseven}
\end{equation}
On the other hand, the quantum field~$V_Q$ transforms as the adjoint
representation under the background transformation and the covariant
derivative for the adjoint representation is defined as
\begin{equation}
   {\cal D}_\alpha V=[\nabla_\alpha,V\},
   \qquad{\cal D}_mV=[\nabla_m,V],
\label{eq:twoxeight}
\end{equation}
where a(n) (anti-)commutator is used when $V$~is
Grassmann-even(-odd). Expressions become even simpler with use of the
adjoint gauge representation matrix, which is defined by
\begin{equation}
   ({\cal T}^a)^{bc}=-it^{abc},
   \qquad\tr{\cal T}^a{\cal T}^b=C_2(G)\delta^{ab}.
\label{eq:twoxnine}
\end{equation}
With this convention, the covariant derivative in the adjoint
representation~(\ref{eq:twoxeight}) can be written as
\begin{equation}
   {\cal D}_\alpha V=T^a(\widetilde\nabla_\alpha V)^a,
   \qquad{\cal D}_m V=T^a(\widetilde\nabla_m V)^a,
\label{eq:twoxten}
\end{equation}
where a component of the covariant derivative is defined by
\begin{equation}
   (\widetilde\nabla_\alpha V)^a
   =(e^{-{\cal V}_B})^{ab}D_\alpha(e^{{\cal V}_B})^{bc}V^c,
   \qquad
   \{\widetilde\nabla_\alpha,\overline D_{\dot\alpha}\}
   =-2i\sigma^m_{\alpha\dot\alpha}\widetilde\nabla_m,
\label{eq:twoxeleven}
\end{equation}
and ${\cal V}_B$~is the background gauge superfield in the adjoint
representation
\begin{equation}
   {\cal V}_B={\cal T}^aV_B^a.
\label{eq:twoxtwelve}
\end{equation}

The essence of the background field method is to use the gauge fixing
condition that is covariant under the background gauge
transformation~(\ref{eq:twoxfive}). We impose the Lorentz-type gauge
fixing condition and its conjugate
\begin{equation}
   \overline D^2V_Q=f,
   \qquad{\cal D}^2V_Q=e^{-V_B}f^\dagger e^{V_B}.
\label{eq:twoxthirteen}
\end{equation}
Then the standard procedure\cite{rf:WEST} gives rise to the gauge
fixing term and the ghost-anti-ghost term:
\begin{eqnarray}
   S'&=&-{\xi\over8T(R)}\int d^8z\,
   \tr(\overline D^2V_Q)({\cal D}^2V_Q)
\nonumber\\
   &&+{1\over T(R)}\int d^8z\,
   \tr(e^{-V_B}c^{\prime\dagger}e^{V_B}+c^\prime)
\nonumber\\
   &&\qquad\qquad\quad\times
   {\cal L}_{V_Q/2}\cdot
   [(c+e^{-V_B}c^\dagger e^{V_B})
    +\coth({\cal L}_{V_Q/2})\cdot
   (c-e^{-V_B}c^\dagger e^{V_B})]
\nonumber\\
   &&-{2\xi\over T(R)}\int d^8z\,\tr e^{-V_B}b^\dagger e^{V_B}b,
\label{eq:twoxfourteen}
\end{eqnarray}
where $\xi$~is the gauge parameter. By construction, this action is
invariant under the background field transformation,
Eq.~(\ref{eq:twoxfive}) and $b'=e^{-i\Lambda}be^{i\Lambda}$ etc. Note
that, since the parameter~$\Lambda$ of the quantum field
transformation~(\ref{eq:twoxfour}) is chiral and the gauge fixing
function~$f$ in~Eq.~(\ref{eq:twoxthirteen}) is also chiral, all the
ghost~$c$, anti-ghost~$c^\prime$, and Nielsen-Kallosh ghost~$b$ are
simply chiral superfields
$\overline D_{\dot\alpha}c=\overline D_{\dot\alpha}c^\prime=%
\overline D_{\dot\alpha}b=0$.

\section{Supersymmetric gauge covariant regularization}
In calculating radiative corrections to the effective action in the
background field method, i.e., the generating functional of 1PI
Green's functions with all the external lines being the background
field $V_B$ or~$\Phi_B$, we expand the total action~$S_T=S+S'$
in powers of the {\it quantum\/} field, $S_T=S_{T0}+S_{T1}+S_{T2}
+S_{T3}+\cdots$. (Hereafter, a number appearing in the subscript
indicates the power of the quantum fields.) The quadratic action is
further decomposed as
$S_{T2}=S_{T2}^{\rm gauge}+S_{T2}^{\rm ghost}+S_{T2}^{\rm chiral}
+S_{T2}^{\rm mix}$.

The first part of the quadratic action, which is composed purely of
the gauge superfields, is given by
\begin{eqnarray}
   &&S_{T2}^{\rm gauge}
\nonumber\\
   &&=\int d^8z\,V_Q^a
   \bigl[-\widetilde\nabla^m\widetilde\nabla_m
   +{1\over2}{\cal W}_B^\alpha\widetilde\nabla_\alpha
   -{1\over2}\overline{\cal W}_{B\dot\alpha}^\prime
    \overline D^{\dot\alpha}
   +{1\over16}(1-\xi)(\widetilde\nabla^2\overline D^2
            +\overline D^2\widetilde\nabla^2)\bigr]^{ab}V_Q^b,
\label{eq:threexone}
\end{eqnarray}
where the field strength in the adjoint representation is defined by
\begin{eqnarray}
   &&{\cal W}_{B\alpha}={\cal T}^aW_{B\alpha}^a
   =-{1\over4}\overline D^2(e^{-{\cal V}_B}D_\alpha e^{{\cal V}_B})
   =-{1\over4}[\overline D_{\dot\alpha},
               \{\overline D^{\dot\alpha},\widetilde\nabla_\alpha\}],
\nonumber\\
   &&\overline{\cal W}_{B\dot\alpha}^\prime
   ={\cal T}^a\overline W_{B\dot\alpha}^{\prime a}
   ={1\over4}e^{-{\cal V}_B}D^2(e^{{\cal V}_B}
   \overline D_{\dot\alpha}e^{-{\cal V}_B})e^{{\cal V}_B}
   ={1\over4}[\widetilde\nabla^\alpha,
              \{\widetilde\nabla_\alpha,\overline D_{\dot\alpha}\}].
\label{eq:threextwo}
\end{eqnarray}
The ghost-anti-ghost action, to second order in the quantum fields,
is given by
\begin{equation}
   S_{T2}^{\rm ghost}
   =\int d^8z\,[c^{\prime\dagger a}(e^{{\cal V}_B})^{ab}c^b
                    +c^{\dagger a}(e^{{\cal V}_B})^{ab}c^{\prime b}
                 -2\xi b^{\dagger a}(e^{{\cal V}_B})^{ab}b^b].
\label{eq:threexthree}
\end{equation}
There are two kinds of action which contain the chiral multiplet. One
is the part that survives even for~$\Phi_B=0$,
\begin{equation}
   S_{T2}^{\rm chiral}=\int d^8z\,\Phi_Q^\dagger e^{V_B}\Phi_Q
   +\int d^6z\,{1\over2}\Phi_Q^Tm\Phi_Q+{\rm h.c.},
\label{eq:threexfour}
\end{equation}
and the other is the part that disappears for~$\Phi_B=0$,
\begin{eqnarray}
   S_{T2}^{\rm mix}
   &=&\int d^8z\,\bigl(\Phi_B^\dagger e^{V_B}V_Q\Phi_Q
                     +\Phi_Q^\dagger e^{V_B}V_Q\Phi_B
            +{1\over2}\Phi_B^\dagger e^{V_B}V_Q^2\Phi_B\bigr)
\nonumber\\
   &&+\int d^6z\,g\Phi_B\Phi_Q^2+{\rm h.c.}
\label{eq:threexfive}
\end{eqnarray}

Our regularization is then implemented as follows: We take propagators
of the quantum fields that are given by formally diagonalizing
$S_2=S_{T2}^{\rm gauge}+S_{T2}^{\rm ghost}+%
S_{T2}^{\rm chiral}$. Then, for a finite ultraviolet cutoff~$M$, we
modify the propagators so as to improve the ultraviolet behavior and
simultaneously preserve the background gauge covariance. For example,
for the quantum gauge superfield, we use\footnote{%
$\delta(z)=\delta(x)\delta(\theta)\delta(\overline\theta)$.}
\begin{eqnarray}
   &&\VEV{{\rm T}^*V_Q^a(z)V_Q^b(z')}
\nonumber\\
   &&={i\over2}\biggl[
   f\bigl([-\widetilde\nabla^m\widetilde\nabla_m
   +{\cal W}_B^\alpha\widetilde\nabla_\alpha/2
   -\overline{\cal W}_{B\dot\alpha}^\prime
    \overline D^{\dot\alpha}/2
   +(1-\xi)(\widetilde\nabla^2\overline D^2
            +\overline D^2\widetilde\nabla^2)/16]
   \bigm/(\xi M^2)\bigr)
\nonumber\\
   &&\qquad\qquad
   \times{1\over-\widetilde\nabla^m\widetilde\nabla_m
   +{\cal W}_B^\alpha\widetilde\nabla_\alpha/2
   -\overline{\cal W}_{B\dot\alpha}^\prime
   \overline D^{\dot\alpha}/2
   +(1-\xi)(\widetilde\nabla^2\overline D^2
            +\overline D^2\widetilde\nabla^2)/16}
   \biggr]^{ab}
\nonumber\\
   &&\qquad\qquad\qquad\qquad\qquad\qquad\qquad\qquad\qquad
   \qquad\qquad\qquad\qquad\qquad\qquad
   \times\delta(z-z'),
\label{eq:threexsix}
\end{eqnarray}
and, for the ghost superfields,
\begin{eqnarray}
   &&\VEV{{\rm T}^*c^a(z)c^{\prime\dagger b}(z')}
   =\VEV{{\rm T}^*c^{\prime a}(z)c^{\dagger b}(z')}
   =-2\xi\VEV{{\rm T}^*b^a(z)b^{\dagger b}(z')}
\nonumber\\
   &&=
   i\bigl[
   f(-\overline D^2\widetilde\nabla^2/16 M^2)
   \overline D^2{1\over\widetilde\nabla^2\overline D^2}
   \widetilde\nabla^2e^{-{\cal V}_B}\bigr]^{ab}
   \delta(z-z'),
\label{eq:threexseven}
\end{eqnarray}
and, for the quantum chiral superfield,
\begin{equation}
   \VEV{{\rm T}^*\Phi_Q(z)\Phi_Q^\dagger(z')}
   ={i\over16}f(-\overline D^2\nabla^2/16 M^2)
   \overline D^2
   {1\over\nabla^2\overline D^2/16-m^\dagger m}\nabla^2e^{-V_B}
   \delta(z-z'),
\label{eq:threexeight}
\end{equation}
and
\begin{equation}
   \VEV{{\rm T}^*\Phi_Q(z)\Phi_Q^T(z')}
   ={i\over4}f(-\overline D^2\nabla^2/16M^2)
   \overline D^2
   {1\over\nabla^2\overline D^2/16-m^\dagger m}m^\dagger
   \delta(z-z').
\label{eq:threexnine}
\end{equation}
In these expressions, $f(t)$~is the regulator, which decreases
sufficiently rapidly $f(\infty)=f'(\infty)=f''(\infty)=\cdots=0$ in
the ultraviolet, and $f(0)=1$ to reproduce the original propagators
in the infinite cutoff limit~$M\to\infty$. In this way, the
propagators obey the same transformation law as the original ones
under the background gauge transformation on the background gauge
superfield~$V_B$. In fact, it can easily be seen that the propagator
of the vector superfield, even with the
modification~(\ref{eq:threexsix}), transforms covariantly under the
background gauge transformation~(\ref{eq:twoxfive}):
\begin{equation}
   \VEV{{\rm T}^*V_Q^a(z)V_Q^b(z')}'
   =\bigl[e^{-i\widetilde\Lambda(z)}\bigr]^{ac}
   \VEV{{\rm T}^*V_Q^c(z)V_Q^d(z')}
   \bigl[e^{i\widetilde\Lambda(z')}\bigr]^{db},
\label{eq:threexten}
\end{equation}
where $\widetilde\Lambda={\cal T}^a\Lambda^a$~is the gauge
parameter in the adjoint representation. Similarly, the
propagator~(\ref{eq:threexeight}) transforms as
\begin{equation}
   \VEV{{\rm T}^*\Phi_Q(z)\Phi_Q^\dagger(z')}'
   =e^{-i\Lambda(z)}\VEV{{\rm T}^*\Phi_Q(z)\Phi_Q^\dagger(z')}
   e^{i\Lambda^\dagger(z')},
\label{eq:threexeleven}
\end{equation}
under the background gauge transformation. These properties are
crucial for the gauge covariance of the scheme.

Using the above propagators of the quantum fields, 1PI Green's
functions are evaluated as follows. There are two kinds of
contributions, because we have diagonalized
$S_2=S_{T2}^{\rm gauge}+S_{T2}^{\rm ghost}+S_{T2}^{\rm chiral}$ in
constructing the propagators. (I)~Most of radiative corrections are
evaluated (as usual) by simply connecting quantum fields
in~$S_{T2}^{\rm mix}$, $S_{T3}$, $S_{T4}$, etc., by the modified
propagators. This defines the first part of the effective action,
${\mit\Gamma}_{\rm I}[V_B,\Phi_B]$, which is given by the 1PI part of
\begin{equation}
   \VEV{{\rm T}^*\exp[i(S_T-S_2)]}.
\label{eq:threextwelve}
\end{equation}
(II)~However, since the quadratic action~$S_2$ depends on the
background gauge superfield~$V_B$ (but not on~$\Phi_B$)
non-trivially, the one-loop determinant arising from the Gaussian
integration of~$\exp(iS_2)$ has to be taken into account. We
{\it define\/} this one-loop effective action as\cite{rf:HAYA,rf:OHS}
\begin{eqnarray}
   {\mit\Gamma}_{\rm II}[V_B]
   &=&\int_0^1dg\,\int d^8z\,V_B^a(z)
   {\delta{\mit\Gamma}_{\rm II}[gV]\over\delta gV_B^a(z)}
\nonumber\\
   &=&\int_0^1dg\,\int d^8z\,V_B^a(z)
   \VEV{{\delta S_2\over\delta V_B^a(z)}}_{V_B\to gV_B}.
\label{eq:threexthirteen}
\end{eqnarray}
Here the indication~$V_B\to gV_B$ implies that all $V_B$-dependences
involved are replaced by~$gV_B$. The quantum fields in the vacuum
expectation value are connected by the modified propagators. This
prescription for the one-loop effective action is characteristic of
our scheme and the prescription maximally respects the gauge
covariance.\cite{rf:HAY,rf:OHS} \ We will discuss this point in more
detail in the next section.

The total effective action is then given by the sum
${\mit \Gamma}[V_B,\Phi_B]=%
{\mit \Gamma}_{\rm I}[V_B,\Phi_B]+{\mit \Gamma}_{\rm II}[V_B]$.

(III)~When a certain composite operator~$O(z)$ is inserted in a
Green's function, it is computed as usual (by using the modified
propagators):
\begin{equation}
   \VEV{{\rm T}^*O(z)\exp[i(S_T-S_2)]}.
\label{eq:threexfourteen}
\end{equation}

It is easy to see that the first part of the effective
action~${\mit\Gamma}_{\rm I}[V_B,\Phi_B]$ (which contains all the
higher loop diagrams) is always supersymmetric and background gauge
invariant. Also a composite operator which behaves classically as a
gauge covariant superfield is regularized as a background gauge
covariant superfield. On the other hand, the one-loop effective
action~${\mit\Gamma}_{\rm II}[V_B]$ defined by
Eq.~(\ref{eq:threexthirteen}) is not necessarily gauge invariant.
This is assuring because if the whole effective action were always
gauge invariant then there would be no possibility of the gauge
anomaly that may arise from chiral multiplet's loop. However,
${\mit\Gamma}_{\rm II}[V_B]$ {\it automatically\/} restores the gauge
invariance {\it without\/} supplementing counterterms, when the gauge
anomaly~$d^{abc}=\tr T^a\{T^b,T^c\}$ vanishes. We will explain this
mechanism in the next section. The
prescription~(\ref{eq:threexfourteen}), on the other hand, is quite
useful in evaluating the quantum anomaly while preserving the
supersymmetry and the gauge covariance (or invariance). We will
present several applications in the later section.

\section{Peculiarity of the one-loop effective action}
Our prescription~(\ref{eq:threexthirteen}) is a natural supersymmetric
generalization of the gauge covariant regularization
in~Ref.~\citen{rf:BAN}. The idea behind Eq.~(\ref{eq:threexthirteen})
is the following. We first introduce an auxiliary gauge coupling
parameter $g$ by~$V_B\to gV_B$. Then we may differentiate the
effective action~${\mit\Gamma}_{\rm II}[gV_B]$ with respect to the
parameter~$g$ and integrate it over this parameter. Noting that the
$g$-dependences arise only through the combination~$gV_B$, we have
the first line of~Eq.~(\ref{eq:threexthirteen}) as the formal
representation. In the second line, the regularization is specified
with use of the modified propagators in the preceding section. The
crucial point for the property of the prescription is the gauge
{\it covariance\/} of the composite
operator\cite{rf:HAY,rf:HAYA,rf:OHS}
\begin{equation}
   \VEV{{\delta S_2\over\delta V_B^a(z)}}'
   =\int d^8z'\,
   {\delta V_B^b(z')\over\delta V_B^{\prime a}(z)}
   \VEV{{\delta S_2\over\delta V_B^b(z')}},
\label{eq:fourxone}
\end{equation}
where the prime stands for the background gauge
transformation~(\ref{eq:twoxfive}).

To see how the prescription~(\ref{eq:threexthirteen}) works, we
consider the gauge variation of~${\mit\Gamma}_{\rm II}[V_B]$ under the
infinitesimal gauge transformation\cite{rf:WES}
\begin{equation}
   \delta_\Lambda V_B=
   i{\cal L}_{V_B/2}\cdot[(\Lambda+\Lambda^\dagger)
   +\coth({\cal L}_{V_B/2})\cdot(\Lambda-\Lambda^\dagger)],
\label{eq:fourxtwo}
\end{equation}
i.e., the consistent gauge anomaly.\cite{rf:PIG} \ The variation
of~Eq.~(\ref{eq:threexthirteen}) yields
\begin{eqnarray}
   \delta_\Lambda{\mit\Gamma}_{\rm II}[V_B]
   &=&\int_0^1dg\,\int d^8z\,\delta_\Lambda V_B^a(z)
   \VEV{{\delta S_2\over\delta V_B^a(z)}}_{V_B\to gV_B}
\nonumber\\
   &&+\int_0^1dg\,\int d^8z\,\int d^8z'\,V_B^a(z)
   \delta_\Lambda V_B^b(z'){\delta\over\delta V_B^b(z')}
   \VEV{{\delta S_2\over\delta V_B^a(z)}}_{V_B\to gV_B}.
\label{eq:fourxthree}
\end{eqnarray}
We then insert~$dg/dg=1$ into the first term and perform the
integration by parts with respect to~$g$. By noting again that the
$g$-dependences arise only through the combination~$gV_B$, we have the
following representation
\begin{eqnarray}
   &&\delta_\Lambda{\mit\Gamma}_{\rm II}[V_B]
   =\int d^8z\,\delta_\Lambda V_B^a(z)
   \VEV{{\delta S_2\over\delta V_B^a(z)}}
\nonumber\\
   &&\quad+\int_0^1dg\,\int d^8z'\,\delta_\Lambda V_B^b(z')
\nonumber\\
   &&\qquad\quad\times
   \int d^8z\,\biggl\{V_B^a(z)
   \biggl[{\delta\over\delta V_B^b(z')}
   \VEV{{\delta S_2\over\delta V_B^a(z)}}
   -{\delta\over\delta V_B^a(z)}
   \VEV{{\delta S_2\over\delta V_B^b(z')}}\biggr]\biggr\}
   _{V_B\to gV_B}.
\label{eq:fourxfour}
\end{eqnarray}
This representation shows that our consistent anomaly consists of two
pieces: The first piece is the covariant gauge
anomaly\cite{rf:KON,rf:HAYA}
\begin{eqnarray}
   &&\int d^8z\,\delta_\Lambda V_B^a(z)
   \VEV{{\delta S_2\over\delta V^a(z)}}
\nonumber\\
   &&\Mto-{1\over64\pi^2}\int d^6z\,\tr i\Lambda
   W_B^\alpha W_{B\alpha}
   +{1\over64\pi^2}\int d^6\overline z\,\tr
   e^{-V_B}i\Lambda^\dagger e^{V_B}
   \overline W_{B\dot\alpha}^\prime\overline W_B^{\prime\dot\alpha},
\label{eq:fourxfive}
\end{eqnarray}
where the conjugate of the background field strength has been defined
by
\begin{equation}
   \overline W_{B\dot\alpha}^\prime
   =e^{-V_B}\overline W_{B\dot\alpha}e^{V_B}
   ={1\over4}[\nabla^\alpha,\{\nabla_\alpha,
              \overline D_{\dot\alpha}\}]
   \quad{\rm and}\quad
   \overline W_{B\dot\alpha}
   ={1\over4}D^2(e^{V_B}\overline D_{\dot\alpha}e^{-V_B}).
\label{eq:fourxsix}
\end{equation}
Note that the covariant anomaly~(\ref{eq:fourxfive}) is proportional
to the anomaly~$d^{abc}$, because $\Lambda$, $W_{B\alpha}$,
$\Lambda^\dagger$, and~$\overline W_{B\dot\alpha}^\prime$ are Lie
algebra valued. The second piece in~Eq.~(\ref{eq:fourxfour}), on the
other hand, provides a difference between the consistent anomaly and
the covariant anomaly. The difference is expressed by the functional
rotation of the covariant gauge current
\begin{equation}
   {\delta\over\delta V_B^b(z')}
   \VEV{{\delta S_2\over\delta V_B^a(z)}}
   -{\delta\over\delta V_B^a(z)}
   \VEV{{\delta S_2\over\delta V_B^b(z')}},
\label{eq:fourxseven}
\end{equation}
which represents the non-integrability of the covariant gauge
current. The important point to note here is that the gauge
covariance~(\ref{eq:fourxone}) implies the following property of the
functional rotation:
\begin{eqnarray}
   &&\int d^8z\,\delta_\Lambda V_B^a(z)
   \biggl[{\delta\over\delta V_B^b(z')}
   \VEV{{\delta S_2\over\delta V_B^a(z)}}
   -{\delta\over\delta V_B^a(z)}
   \VEV{{\delta S_2\over\delta V_B^b(z')}}\biggr]
\nonumber\\
   &&={\delta\over\delta V_B^b(z')}
   \int d^8z\,\delta_\Lambda V_B^a(z)
   \VEV{{\delta S_2\over\delta V_B^a(z)}}.
\label{eq:fourxeight}
\end{eqnarray}
The right hand side is nothing but the covariant
anomaly~(\ref{eq:fourxfive}). Quite interestingly, the functional
rotation~(\ref{eq:fourxseven}) is a {\it local\/} functional of the
gauge superfield, being proportional to (a derivative of) the delta
function~$\delta(z-z')$, as we will see shortly. This fact and
Eq.~(\ref{eq:fourxeight}) imply that the functional rotation vanishes
and consequently our consistent anomaly~(\ref{eq:fourxfour}) entirely
vanishes, when the covariant anomaly vanishes. Namely, when the gauge
representation of the chiral multiplet is anomaly-free, i.e.,
when~$d^{abc}=0$, the one-loop effective
action~(\ref{eq:threexthirteen}) automatically restores the gauge
invariance {\it without\/} supplementing any counterterms. In this
sense, a breaking of the gauge symmetry is kept to be minimal with
the present prescription.

The direct calculation shows that for arbitrary variations~$\delta_1$
and~$\delta_2$,\cite{rf:OHS}
\begin{eqnarray}
   &&\delta_1\VEV{\delta_2S_2}-\delta_2\VEV{\delta_1S_2}
\nonumber\\
   &&\Mto{1\over64\pi^2}\int d^8z\,
   \tr\Delta_1\bigl(
   [{\cal D}^\alpha\Delta_2,W_{B\alpha}]
   +[\overline D_{\dot\alpha}\Delta_2,
          \overline W_B^{\prime\dot\alpha}]
   +\{\Delta_2,{\cal D}^\alpha W_{B\alpha}\}\bigr),
\label{eq:fourxnine}
\end{eqnarray}
with the notation~$\Delta=e^{-V_B}\delta e^{V_B}$. From this,
we see that the functional rotation~(\ref{eq:fourxseven}) is a local
object. The consistent gauge anomaly~(\ref{eq:fourxfour}) is
consequently given by\cite{rf:OHS}
\begin{eqnarray}
   &&\delta_\Lambda{\mit\Gamma}_{\rm II}[V_B]
\nonumber\\
   &&\Mto-{1\over64\pi^2}\int d^6z\,\tr i\Lambda
   W_B^\alpha W_{B\alpha}
   +{1\over64\pi^2}\int d^6\overline z\,\tr e^{-V_B}
   i\Lambda^\dagger e^{V_B}
   \overline W_{B\dot\alpha}^\prime\overline W_B^{\prime\dot\alpha}
\nonumber\\
   &&\qquad
   +{1\over64\pi^2}\int d^8z\,
   \int_0^1dg\int_0^1d\beta\,
   \tr e^{-\beta gV_B}\delta_\Lambda V_Be^{\beta gV_B}
\nonumber\\
   &&\qquad\qquad\qquad\qquad\quad
   \times
   \bigl([{\cal D}^\alpha V_B,W_{B\alpha}]
   +[\overline D_{\dot\alpha}V_B,\overline W_B^{\prime\dot\alpha}]
   +\{V_B,{\cal D}^\alpha W_{B\alpha}\}\bigr)_{V_B\to gV_B}.
\label{eq:fourxten}
\end{eqnarray}
It is obvious that our consistent anomaly is proportional to the
anomaly~$d^{abc}$, as expected. This anomaly must satisfy the
Wess-Zumino consistency condition because it is the gauge variation
of the functional~(\ref{eq:threexthirteen}). In fact,
Eq.~(\ref{eq:fourxten}) coincides the consistent anomaly due to
Marinkovi\'c,\cite{rf:PIG} which was obtained as a solution of the
consistency condition.

We are interested in anomaly-free models for which the gauge
anomaly~(\ref{eq:fourxten}) vanishes. Nevertheless it is interesting
to examine the form of the anomaly~(\ref{eq:fourxten}) in the
Wess-Zumino~(WZ) gauge.\cite{rf:WES} \ We first set\footnote{%
$y^m=x^m+i\theta\sigma^m\overline\theta$.} $\Lambda(z)=a(y)$ to
reproduce the usual gauge transformation ($a$~is real). Then we have
\begin{eqnarray}
   \delta_\Lambda{\mit\Gamma}_{\rm II}[V_B]
   &\Mto&-{1\over96\pi^2}\int d^4x\,\tr a
   \bigl[\varepsilon^{mnkl}\partial_m
   \bigl(v_{Bn}\partial_kv_{Bl}+{i\over4}v_{Bn}v_{Bk}v_{Bl}\bigr)
\nonumber\\
   &&\qquad\qquad\qquad\qquad\quad
   -{1\over2}\partial_m(\overline\lambda_B\overline\sigma^m\lambda_B
   -\lambda_B\sigma^m\overline\lambda_B)\bigr].
\label{eq:fourxeleven}
\end{eqnarray}
This expression of the usual gauge anomaly in the WZ gauge is
surprisingly simple compared to the result of existing field
theoretical calculations. The first term is Bardeen's anomaly; the
second term may be eliminated by adding a non-super\-symmetric local
counterterm~$C$
\begin{equation}
   C={1\over384\pi^2}\int d^4x\,\tr
   v_B^m(\overline\lambda_B\overline\sigma_m\lambda_B
       -\lambda_B\sigma_m\overline\lambda_B).
\label{eq:fourxtwelve}
\end{equation}
as~$\delta_\Lambda{\mit\Gamma}_{\rm II}[V_B]+%
\delta_aC$, where $\delta_a$ is the usual gauge transformation.

As another interesting case, we may consider the anomalous breaking
of the supersymmetry in the WZ gauge, the so-called
$\varepsilon$-SUSY anomaly.\cite{rf:ITO,rf:GUA} \
The super-transformation in the WZ gauge is a combination of the
supersymmetric transformation
generated by~$\varepsilon Q+\overline\varepsilon\overline Q$ (which
is not anomalous in the present formulation) {\it and\/}
the gauge transformation~$\delta_\Lambda$ with the gauge
parameter~$\Lambda(z)=-i\theta\sigma^m\overline\varepsilon v_{Bm}(y)-%
\theta^2\overline\varepsilon\overline\lambda_B(y)$.\cite{rf:WES} \ 
Therefore we have the (apparent) breaking of supersymmetry as the
consequence of the gauge anomaly. By setting the gauge
parameter~$\Lambda$ to this form in~Eq.~(\ref{eq:fourxten}), we have
\begin{eqnarray}
   \delta_\Lambda{\mit\Gamma}_{\rm II}[V]
   &\Mto&{i\over384\pi^2}\int d^4x\,
   \tr(\overline\varepsilon\overline\sigma^m\lambda_B
   -\overline\lambda_B\overline\sigma^m\varepsilon)
\nonumber\\
   &&
   \times\bigl\{
   3\overline\lambda_B\overline\sigma_m\lambda_B
   -\varepsilon_m{}^{nkl}\bigl[2v_{Bn}(\partial_kv_{Bl})
   +2(\partial_nv_{Bk})v_{Bl}
   +{3i\over2}v_{Bn}v_{Bk}v_{Bl}\bigr]\bigr\}
\nonumber\\
   &&-\delta_\varepsilon C,
\label{eq:fourxthirteen}
\end{eqnarray}
where $\delta_\varepsilon$~is the super-transformation in the WZ
gauge $\delta_\varepsilon v_B^m=%
i\overline\varepsilon\overline\sigma^m\lambda_B+{\rm h.c.}$,
$\delta_\varepsilon\lambda_B=%
\sigma^{mn}\varepsilon v_{Bmn}+i\varepsilon D_B$.
Eq.~(\ref{eq:fourxthirteen}) shows that Eq.~(\ref{eq:fourxten})
reproduces the $\varepsilon$-SUSY anomaly given
in~Ref.~\citen{rf:GUA} again with the non-supersymmetric local
counterterm~$C$~(\ref{eq:fourxtwelve}). The structure
of~Eq.~(\ref{eq:fourxtwelve}) is quite simple, compared to that of
the counterterm required in~Ref.~\citen{rf:GUA} for obtaining the
above ``minimal'' form. Our anomaly is proportional to~$d^{abc}$ from
the beginning and thus the possible (non-supersymmetric) counterterm
also must be proportional to~$d^{abc}$. This fact severely restricts
the possible form of (non-supersymmetric) counterterms.

For anomaly-free cases, the above prescription for the one-loop
effective action~(\ref{eq:threexthirteen}) can be shown to be
equivalent\cite{rf:HAYA} to (a supersymmetric generalization of) the
generalized Pauli-Villars regularization in~Ref.~\citen{rf:FRO}.
Since this is a Lagrangian level regularization whose Hamiltonian is
Hermitian, the S-matrix is manifestly unitary. (Note that, in the
$M\to\infty$~limit, negative norm Pauli-Villars regulators cannot
contribute to the out-state of the physical S-matrix.)

\section{One-loop Green's functions}
It is easy to evaluate one-loop two point Green's functions with our
scheme.\cite{rf:HAY} \ To carry out actual calculations, we have to
choose a form of the regulator. A simple choice is
\begin{equation}
   f(t)={1\over(t+1)^2}.
\label{eq:fivexone}
\end{equation}
The self-energy part of the chiral multiplet, for example, is
evaluated according to the rule~(\ref{eq:threextwelve}). This yields
in terms of the effective action,\footnote{%
For simplicity, we neglect the effect of the superpotential by
setting~$m=g=0$.}
\begin{eqnarray}
   {\mit\Gamma}_{\rm I}[V_B=0,\Phi_B]
   &=&-{1\over32\pi^2}C(R)
   \int d^4\theta\int d^4x\,d^4x'\,
   \Phi_B^\dagger(x,\theta,\overline\theta)
   \Phi_B(x',\theta,\overline\theta)
\nonumber\\
   &&\qquad\qquad\qquad
   \times\int{d^4p\over(2\pi)^4}\,e^{ip(x-x')}
   \Bigl(\ln{M^2\over p^2}-{5\over6}\Bigr)+O(\Phi_B).
\label{eq:fivextwo}
\end{eqnarray}
The constant~$-5/6$ depends on the choice of the regulator~$f(t)$. \
In fact, since we ``know'' that the effective
action~${\mit\Gamma}_{\rm I}[V_B,\Phi_B]$ is always gauge invariant,
we may covariantize the local part of the effective action (which is
proportional to~$\ln M^2$)
as~$\int d^8z\,\Phi_B^\dagger e^{V_B}\Phi_B$.

For the vacuum polarization tensor of~$V_B$, we have from the
prescription~(\ref{eq:threexthirteen}),
\begin{eqnarray}
   &&{\mit\Gamma}_{\rm II}[V_B]
   ={M^2\over16\pi^2}\int d^8z\,\tr V_B
   +{1\over64\pi^2}{T(R)-3C_2(G)\over2T(R)}
   \int d^4\theta\int d^4x\,d^4x'
\nonumber\\
   &&\times\tr V_B(x,\theta,\overline\theta)
   \Bigl[{1\over4}D^\alpha\overline D^2D_\alpha
   V_B(x',\theta,\overline\theta)\Bigr]
   \int{d^4p\over(2\pi)^4}\,e^{ip(x-x')}
   \Bigl(\ln{M^2\over p^2}+1\Bigr)+O(V_B^3).
\label{eq:fivexthree}
\end{eqnarray}
For anomaly-free cases, we again know that the effective action is
gauge invariant without counterterms. Thus we may covariantize the
local term proportional to~$\ln M^2$
as~$\int d^6z\,\tr W_B^\alpha W_{B\alpha}$.
Equations~(\ref{eq:fivexthree}) and~(\ref{eq:fivexthree}) reproduces
the well-known one-loop result.\cite{rf:WEST} \ If one is interested
in the divergent part of the effective
action~${\mit\Gamma}_{\rm II}[V_B]$, it is easy to obtain the general
result for an {\it arbitrary\/} $f(t)$:\cite{rf:HAY}
\begin{eqnarray}
   &&M{d\over dM}{\mit\Gamma}_{\rm II}[V_B]
\nonumber\\
   &&\Mto
   {M^2\over8\pi^2}\int_0^\infty dt\,f(t)\int d^8z\,
   \tr V_B
   +{T(R)-3C_2(G)\over64\pi^2}\int d^6z\,
   \tr W_B^\alpha W_{B\alpha}.
\label{eq:fivexfour}
\end{eqnarray}
This shows that the one-loop $\beta$-function of the gauge coupling
is independent of the choice of~$f(t)$.

\section{Super-chiral and superconformal anomalies}
Since our scheme gives a supersymmetric gauge covariant definition of
composite operators, it also provides a simple and reliable method to
evaluate quantum anomalies. In this section, we present several
examples in the one-loop approximation.\footnote{%
Throughout this section, we assume that the background chiral
superfield~$\Phi_B$ and the Yukawa coupling~$g$ vanish, for
simplicity of analysis.}

The first example is the super-chiral
anomaly,\cite{rf:CLA,rf:GAT,rf:KON} which is defined as a breaking of
the Ward-Takahashi identity:
\begin{equation}
   -{1\over4}\overline D^2
   \VEV{\Phi^\dagger e^V\Phi(z)}+\VEV{\Phi^Tm\Phi(z)}=0.
\label{eq:sixxone}
\end{equation}
This identity is associated with the chiral symmetry of the massless
action, $\Phi(z)\rightarrow e^{i\alpha}\Phi(z)$. We first take
in~Eq.~(\ref{eq:sixxone}) the quadratic terms in the quantum fields
(i.e., the one-loop approximation). Then, according
to~Eq.~(\ref{eq:threexfourteen}), we define the regularized
super-chiral current as
\begin{equation}
   \VEV{\Phi_Q^\dagger e^{V_B}\Phi_Q(z)}=
   {i\over16}\lim_{z'\rightarrow z}
   \tr f(-\overline D^2\nabla^2/16M^2)\overline D^2
   {1\over\nabla^2\overline D^2/16-m^\dagger m}\nabla^2\delta(z-z'),
\label{eq:sixxtwo}
\end{equation}
and similarly, from~Eq.~(\ref{eq:threexnine}),
\begin{equation}
   \VEV{\Phi_Q^Tm\Phi_Q(z)}=
   {i\over4}\lim_{z'\rightarrow z}
   \tr f(-\overline D^2\nabla^2/16M^2)\overline D^2
   {1\over\nabla^2\overline D^2/16-m^\dagger m}m^\dagger
   m\delta(z-z').
\label{eq:sixxthree}
\end{equation}
We then directly apply $-\overline D^2/4$ on the composite operator
in~Eq.~(\ref{eq:sixxtwo}). Then, by noting the chirality
of~Eq.~(\ref{eq:sixxtwo}) with respect to the $z$~variable, we have
\begin{eqnarray}
   &&-{1\over4}\overline D^2
   \VEV{\Phi_Q^\dagger e^{V_B}\Phi_Q(z)}+\VEV{\Phi_Q^Tm\Phi_Q(z)}
\nonumber\\
   &&=-{i\over4}\lim_{z'\rightarrow z}
   \tr f(-\overline D^2\nabla^2/16M^2)
   \overline D^2\delta(z-z')
\nonumber\\
   &&\Mto
   -{1\over64\pi^2}\tr W_B^\alpha W_{B\alpha}(z),
\label{eq:sixxfour}
\end{eqnarray}
which reproduces the well-known form of the super-chiral anomaly.
(Note that the anomaly is independent of the choice of~$f(t)$.) For
the actual calculation of the second line, see Ref.~\citen{rf:HAY}.
The expression~(\ref{eq:sixxfour}) holds even in {\it chiral\/} gauge
theories, and in this sense Eq.~(\ref{eq:sixxfour}) may be viewed as
a supersymmetric version of the fermion number anomaly.\cite{rf:THO}

Since we have defined the regularized composite operator
in~Eqs.~(\ref{eq:sixxtwo}) and~(\ref{eq:sixxthree}), an anomalous
supersymmetric commutation relation associated with the super-chiral
anomaly, the Konishi anomaly,\cite{rf:KONI} can be derived
straightforwardly. First we note in the Wess-Zumino
gauge,\cite{rf:WES}
\begin{equation}
   \Phi^\dagger e^V\Phi
   =A^\dagger A
   +\sqrt{2}\,\overline\theta\overline\psi A+\cdots,
\label{eq:sixxfive}
\end{equation}
and thus classically,
\begin{equation}
   \sqrt{2}\,\overline\psi^{\dot\alpha}A
   =\overline D^{\dot\alpha}
   (\Phi^\dagger e^V\Phi)\bigr|_{\theta=\overline\theta=0}.
\label{eq:sixxsix}
\end{equation}
Therefore the supersymmetric transformation of the composite
operator~$\overline\psi^{\dot\alpha}A$ may be defined as
\begin{eqnarray}
   {1\over2\sqrt{2}}
   \VEV{
   \{\overline Q_{\dot\alpha},\overline\psi_Q^{\dot\alpha}A_Q(x)\}}
   &=&{1\over4}\overline Q_{\dot\alpha}\overline D^{\dot\alpha}
   \VEV{
   \Phi_Q^\dagger e^{V_B}\Phi_Q(z)}
   \bigr|_{\theta=\overline\theta=0}
\nonumber\\
   &=&{1\over4}\overline D^2
   \VEV{
   \Phi_Q^\dagger e^{V_B}\Phi_Q(z)}
   \bigr|_{\theta=\overline\theta=0}
\nonumber\\
   &\Mto&\VEV{A_Q^TmA_Q(x)}
   -{1\over64\pi^2}
   \tr\lambda_B^\alpha\lambda_{B\alpha}(x).
\label{eq:sixxseven}
\end{eqnarray}
This is the Konishi anomaly.\cite{rf:KONI} \ The point in the above
derivation is that we have defined the regularized composite operator
first; moreover, we did so in terms of the {\it superfield}.
Therefore the supersymmetric transformation of the regularized
composite operator can be performed by one stroke of the differential
operator, $\overline Q_{\dot\alpha}%
=-\partial/\partial\overline\theta^{\dot\alpha}
+i\theta^\alpha\sigma_{\alpha\dot\alpha}^m\partial_m$.\cite{rf:WES}

Another interesting example is the superconformal
anomaly.\cite{rf:FER,rf:LUK} \ It is a breaking of the Ward-Takahashi
identity:
\begin{equation}
   \overline D^{\dot\alpha}
   \VEV{R_{\alpha\dot\alpha}(z)}
   -2\VEV{\Phi^Tme^{-V}D_\alpha e^V\Phi(z)}
   +{2\over3}D_\alpha\VEV{\Phi^Tm\Phi(z)}=0.
\label{eq:sixxeight}
\end{equation}
The superconformal current~$R_{\alpha\dot\alpha}$ is defined by
$R_{\alpha\dot\alpha}=R_{\alpha\dot\alpha}^{\rm chiral}
+R_{\alpha\dot\alpha}^{\rm gauge}$, where
\begin{eqnarray}
   R_{\alpha\dot\alpha}^{\rm chiral}
   &=&
   -\overline D_{\dot\alpha}(\Phi^\dagger e^V)e^{-V}D_\alpha e^V\Phi
   -{1\over3}[D_\alpha,\overline D_{\dot\alpha}]
   (\Phi^\dagger e^V\Phi)
\nonumber\\
   &=&-\overline D_{\dot\alpha}(\Phi_Q^\dagger e^{V_B})
   \nabla_\alpha\Phi_Q
   -{1\over3}[D_\alpha,\overline D_{\dot\alpha}]
   (\Phi_Q^\dagger e^{V_B}\Phi_Q)
   +\cdots,
\label{eq:sixxnine}
\end{eqnarray}
and
\begin{equation}
   R_{\alpha\dot\alpha}^{\rm gauge}
   =-{2\over T(R)}
   \tr W_\alpha e^{-V}\overline W_{\dot\alpha}e^V.
\label{eq:sixxten}
\end{equation}

From~(\ref{eq:sixxnine}), the regularized superconformal current of
the chiral multiplet is defined to the one-loop level by
\begin{eqnarray}
   &&\VEV{R_{\alpha\dot\alpha}^{\rm chiral}(z)}
\nonumber\\
   &&=-{i\over16}\lim_{z'\rightarrow z}\tr
   \nabla_\alpha f(-\overline D^2\nabla^2/16M^2)
   \overline D^2{1\over\nabla^2\overline D^2/16-m^\dagger m}\nabla^2
   \overline D_{\dot\alpha}\delta(z-z')
\nonumber\\
   &&\quad-{1\over3}{i\over16}[D_\alpha,\overline D_{\dot\alpha}]
   \lim_{z'\rightarrow z}\tr
   f(-\overline D^2\nabla^2/16M^2)
   \overline D^2{1\over\nabla^2\overline D^2/16-m^\dagger m}\nabla^2
   \delta(z-z').
\label{eq:sixxeleven}
\end{eqnarray}
To deal with this expression, we note the identity
\begin{equation}
   D_\alpha\lim_{z'\rightarrow z}\tr A(z)\delta(z-z')
   =\lim_{z'\rightarrow z}
   \tr[\nabla_\alpha,A(z)\}\delta(z-z'),
\label{eq:sixxtwelve}
\end{equation}
where $A(z)$~is an arbitrary operator. Then by using this identity
and noting the chirality, we find
\begin{eqnarray}
   &&\VEV{R_{\alpha\dot\alpha}^{\rm chiral}(z)}
\nonumber\\
   &&={1\over3}{i\over16}\lim_{z'\rightarrow z}\tr
   \Bigl[
   -\nabla_\alpha f(-\overline D^2\nabla^2/16M^2)
   \overline D^2{1\over\nabla^2\overline D^2/16-m^\dagger m}\nabla^2
   \overline D_{\dot\alpha}
\nonumber\\
   &&\qquad\qquad\qquad\quad
   +f(-\overline D^2\nabla^2/16M^2)
   \overline D^2{1\over\nabla^2\overline D^2/16-m^\dagger m}\nabla^2
   \overline D_{\dot\alpha}\nabla_\alpha
\nonumber\\
   &&\qquad\qquad\qquad\quad
   +\overline D_{\dot\alpha}\nabla_\alpha
   f(-\overline D^2\nabla^2/16M^2)
   \overline D^2{1\over\nabla^2\overline D^2/16-m^\dagger m}\nabla^2
   \Bigr]\delta(z-z').
\label{eq:sixxthirteen}
\end{eqnarray}
By applying~$\overline D^{\dot\alpha}$ further, we have
\begin{eqnarray}
   &&\overline D^{\dot\alpha}
   \VEV{R_{\alpha\dot\alpha}^{\rm chiral}(z)}
\nonumber\\
   &&={1\over3}{i\over16}\lim_{z'\rightarrow z}\tr
   \Bigl[
   \nabla_\alpha f(-\overline D^2\nabla^2/16M^2)
   \overline D^2{1\over\nabla^2\overline D^2/16-m^\dagger m}\nabla^2
   \overline D^2
\nonumber\\
   &&\qquad\qquad\qquad\quad
   -f(-\overline D^2\nabla^2/16M^2)
   \overline D^2{1\over\nabla^2\overline D^2/16-m^\dagger m}\nabla^2
   \overline D_{\dot\alpha}\nabla_\alpha\overline D^{\dot\alpha}
\nonumber\\
   &&\qquad\qquad\qquad\quad
   -\overline D^2\nabla_\alpha
   f(-\overline D^2\nabla^2/16M^2)
   \overline D^2{1\over\nabla^2\overline D^2/16-m^\dagger m}\nabla^2
   \Bigr]\delta(z-z').
\label{eq:sixxfourteen}
\end{eqnarray}
Then we use identities
\begin{eqnarray}
   &&\nabla^2\overline D_{\dot\alpha}\nabla_\alpha
   \overline D^{\dot\alpha}
   =-{1\over2}\nabla^2\overline D^2\nabla_\alpha
   -2\nabla^2 W_{B\alpha},
\nonumber\\
   &&\overline D^2\nabla_\alpha\overline D^2
   =-4W_{B\alpha}\overline D^2,
\label{eq:sixxfifteen}
\end{eqnarray}
to yield
\begin{eqnarray}
   &&\overline D^{\dot\alpha}
   \VEV{R_{\alpha\dot\alpha}^{\rm chiral}(z)}
   -2\VEV{\Phi_Q^Tm\nabla_\alpha\Phi_Q(z)}
   +{2\over3}D_\alpha\VEV{\Phi_Q^Tm\Phi_Q(z)}
\nonumber\\
   &&={i\over2}\lim_{z'\rightarrow z}\tr
   \nabla_\alpha f(-\overline D^2\nabla^2/16M^2)
   \overline D^2\delta(z-z')
\nonumber\\
   &&\quad-{i\over6}D_\alpha
   \lim_{z'\rightarrow z}\tr
   f(-\overline D^2\nabla^2/16M^2)
   \overline D^2\delta(z-z')
\nonumber\\
   &&\quad+2{i\over16}\lim_{z'\rightarrow z}\tr W_{B\alpha}
   f(-\overline D^2\nabla^2/16M^2)
   \overline D^2{1\over\nabla^2\overline D^2/16-m^\dagger m}\nabla^2
   \delta(z-z'),
\label{eq:sixxsixteen}
\end{eqnarray}
where we have used Eq.~(\ref{eq:sixxtwelve}) again. In this
expression, the first two lines on the right-hand side are regarded
as the anomaly, and the last line can be interpreted as a composite
operator. In this way, we obtain\cite{rf:HAY}
\begin{eqnarray}
   &&\overline D^{\dot\alpha}
   \VEV{R_{\alpha\dot\alpha}^{\rm chiral}(z)}
   -2\VEV{\Phi_Q^Tm\nabla_\alpha\Phi_Q(z)}
   +{2\over3}D_\alpha\VEV{\Phi_Q^Tm\Phi_Q(z)}
\nonumber\\
   &&\Mto
   -{1\over8\pi^2}
   \bigl[M^2\int_0^\infty dt\,f(t)+{1\over6}\dA\,\bigr]
   \tr W_{B\alpha}(z)
   +{1\over192\pi^2}D_\alpha\tr W_B^\beta W_{B\beta}(z)
\nonumber\\
   &&\qquad+2
   \tr W_{B\alpha}(z)
   \VEV{\Phi_Q(z)\Phi_Q^\dagger(z)}
   e^{V_B(z)}.
\label{eq:sixxseventeen}
\end{eqnarray}
For the gauge sector, a similar calculation shows\cite{rf:HAYA}
\begin{equation}
   \overline D^{\dot\alpha}\VEV{R_{\alpha\dot\alpha}^{\rm gauge}(z)}
   \Mto
   -2\tr W_{B\alpha}(z)\VEV{\Phi_Q(z)\Phi_Q^\dagger(z)}e^{V_B(z)}
   -{1\over64\pi^2}
   {C_2(G)\over T(R)}D_\alpha\tr W_B^\beta W_{B\beta},
\label{eq:sixxeighteen}
\end{equation}
up to the BRS exact piece which can be neglected due to the
Slavnov-Taylor identity.\cite{rf:HAYA} \ Finally, we have
\begin{eqnarray}
   &&\overline D^{\dot\alpha}\VEV{R_{\alpha\dot\alpha}(z)}
   -2\VEV{\Phi_Q^Tm\nabla_\alpha\Phi_Q(z)}
   +{2\over3}D_\alpha\VEV{\Phi_Q^Tm\Phi_Q(z)}
\nonumber\\
   &&\Mto
   -{1\over8\pi^2}
   \bigl[M^2\int_0^\infty dt\,f(t)+{1\over6}\dA\,\bigr]
   \tr W_{B\alpha}(z)
   -{3C_2(G)-T(R)\over 192\pi^2T(R)}
   D_\alpha\tr W_B^\beta W_{B\beta},
\label{eq:sixxnineteen}
\end{eqnarray}
where the first divergent piece can be removed by adding the
Fayet-Iliopoulos $D$-term.

As we have done for the Konishi anomaly, we may derive from the
superconformal anomaly~(\ref{eq:sixxnineteen}) the anomalous
``central extension'' of the $N=\nobreak1$~supersymmetry algebra
which has recently been advocated by Shifman et.~al.\cite{rf:DVA} \
An analysis of this problem from the viewpoint of path integrals and
the Bjorken-Johnson-Low prescription can be found
in~Ref.~\citen{rf:FUJI}. We first note the definition of the
supercharge:
\begin{equation}
   \overline Q_{\dot\alpha}=\int d^3x\,\overline J_{\dot\alpha}^0,
   \qquad
   \overline J_{\dot\alpha}^m
   =-{1\over2}\overline\sigma^{m\dot\beta\beta}
   \overline J_{\dot\alpha\dot\beta\beta}.
\label{eq:sixxtwenty}
\end{equation}
The improved supercurrent~$\overline J_{\dot\beta\dot\alpha\alpha}$
is related to the superconformal current~$R_{\alpha\dot\alpha}$
as\cite{rf:FER}
\begin{equation}
   R_{\alpha\dot\alpha}=
   R^{(0)}_{\alpha\dot\alpha}
   -i\overline\theta_{\dot\beta}
   \bigl(\overline J^{\dot\beta}{}_{\dot\alpha\alpha}
   -{2\over3}\delta_{\dot\alpha}^{\dot\beta}
   \overline J^{\dot\gamma}{}_{\dot\gamma\alpha}\bigr)+\cdots,
\label{eq:sixxtwentyone}
\end{equation}
where the first component~$R^{(0)}_{\alpha\dot\alpha}$ is the
$R$-current. From these relations, we have classically
\begin{equation}
   \overline J_{\dot\alpha}^0
   =-{i\over2}\overline\sigma^{0\dot\beta\beta}
   (\overline D_{\dot\alpha}R_{\beta\dot\beta}
   -2\varepsilon_{\dot\alpha\dot\beta}\overline D^{\dot\gamma}
   R_{\beta\dot\gamma})\bigr|_{\theta=\overline\theta=0}.
\label{eq:sixxtwentytwo}
\end{equation}
Therefore we may define the supersymmetric transformation of the
supercurrent operator as
\begin{eqnarray}
   \VEV{\{\overline Q_{\dot\alpha},\overline J_{\dot\beta}^0(x)\}}
   &=&
   -{i\over2}\overline\sigma^{0\dot\gamma\gamma}
   \overline Q_{\dot\alpha}
   \bigl[\overline D_{\dot\beta}
   \VEV{R_{\gamma\dot\gamma}(z)}
   -2\varepsilon_{\dot\beta\dot\gamma}\overline D^{\dot\delta}
   \VEV{R_{\gamma\dot\delta}(z)}
   \bigr]\bigr|_{\theta=\overline\theta=0}
\nonumber\\
   &=&{i\over2}
   \overline\sigma^{0\dot\gamma\gamma}
   (\varepsilon_{\dot\alpha\dot\beta}\overline D_{\dot\gamma}
   +2\varepsilon_{\dot\beta\dot\gamma}\overline D_{\dot\alpha})
   \overline D^{\dot\delta}
   \VEV{R_{\gamma\dot\delta}(z)}
   \bigr|_{\theta=\overline\theta=0}
\nonumber\\
   &\Mto&{1\over48\pi^2}{3C_2(G)-T(R)\over T(R)}
   \overline\sigma^{0i}_{\dot\alpha\dot\beta}\partial_i
   \tr\lambda_B^\alpha\lambda_{B\alpha}(x).
\label{eq:sixxtwentythree}
\end{eqnarray}
where we have used the superconformal
anomaly~(\ref{eq:sixxnineteen}).\footnote{%
We have set~$m=0$ for simplicity.}
The ``central extension'' of the $N=1$~supersymmetry
algebra\cite{rf:DVA} is obtained by
integrating Eq.~(\ref{eq:sixxtwentythree}) over the spatial
coordinate~$x$. In deriving the second line from the first line
in~Eq.~(\ref{eq:sixxtwentythree}), we have used the identity
\begin{equation}
   \overline D_{\dot\alpha}\overline D_{\dot\beta}
   X_{\gamma\dot\gamma}\bigr|_{\theta=\overline\theta=0}
   =-\varepsilon_{\dot\alpha\dot\beta}
   \overline D_{\dot\gamma}\overline D^{\dot\delta}
   X_{\gamma\dot\delta}\bigr|_{\theta=\overline\theta=0},
\label{eq:sixxtwentyfour}
\end{equation}
which may be confirmed component by component.

In a similar way, we may study the anomalous commutator between the
supercharge and the $R$-current.\cite{rf:ITOY} \ The $R$-current is
defined from the superconformal current by
\begin{equation}
   R^{(0)m}
   =-{1\over2}\overline\sigma^{m\dot\alpha\alpha}
   R_{\alpha\dot\alpha}\bigr|_{\theta=\overline\theta=0}.
\label{eq:sixxtwentyfive}
\end{equation}
Then the supersymmetric transformation of the $R$-current is related
to the superconformal anomaly as
\begin{eqnarray}
   &&\VEV{[\overline Q_{\dot\alpha},R^{(0)0}(x)]}
   =-{1\over2}\overline\sigma^{0\dot\beta\beta}
   \overline Q_{\dot\alpha}
   \VEV{R_{\beta\dot\beta}(z)}
   \bigr|_{\theta=\overline\theta=0}
\nonumber\\
   &&=-i\VEV{\overline J_{\dot\alpha}^0(x)}
   -\overline\sigma^{0\dot\beta\beta}
   \varepsilon_{\dot\alpha\dot\beta}
   \overline D^{\dot\gamma}
   \VEV{R_{\beta\dot\gamma}(z)}
   \bigr|_{\theta=\overline\theta=0}
\nonumber\\
   &&=-i\VEV{\overline J_{\dot\alpha}^0(x)}
   +{i\over8\pi^2}
   \bigl[M^2\int_0^\infty dt\,f(t)+{1\over6}\dA\,\bigr]
   \tr\lambda_B^\alpha(x)\sigma_{\alpha\dot\alpha}^0
\nonumber\\
   &&\qquad
   +{i\over48\pi^2}{3C_2(G)-T(R)\over T(R)}
   \tr\lambda_B^\alpha(x)
   \Bigl[{1\over2}\sigma_{\alpha\dot\alpha}^0D_B(x)
   +\sigma_{\alpha\dot\alpha}^mv_B^{+0}{}_m(x)\Bigr],
\label{eq:sixxtwentysix}
\end{eqnarray}
where we have used the relation~(\ref{eq:sixxtwentytwo}) and the
superconformal anomaly~(\ref{eq:sixxnineteen}).\footnote{%
We have again set~$m=0$ for simplicity.}
In the last line, $v_{Bmn}^+$~is the self-dual part of the field
strength:
\begin{equation}
   v_{Bmn}^+=
   {1\over2}\bigl(v_{Bmn}+{i\over2}\varepsilon_{mnkl}v_B^{kl}\bigr).
\label{eq:sixxtwentyseven}
\end{equation}

\section{Conclusion}
In this talk, I introduced the recent attempt of our group for
obtaining an invariant regularization scheme in supersymmetric
chiral gauge theories. I also presented several applications in the
one-loop level. Although our scheme provides a unified treatment of
various problems in the one-loop level, the application to
higher-loop problems seems practically difficult. One reason is that
use of the regulator~$f(t)$ introduces complication in the Feynman
rule. Another reason is the lack of the manifest BRS symmetry.
Nevertheless, we believe the idea to use the covariant gauge current,
such as in~Eq.~(\ref{eq:threexthirteen}), is basically the correct
one. We are thus supposing that a certain variation of the dimensional
reduction\cite{rf:SIE} along the above idea might make the scheme 
more practical.

\section*{Acknowledgements}
I would like to thank the organizers of this workshop especially
Professor Takeo Inami and Professor Ryu Sasaki who recommended me
to give a talk on our work.

\end{document}